\documentclass[journal]{IEEEtran}
\usepackage{graphicx}
\usepackage{subfigure}
\usepackage{epsfig}

\begin{document}
\title{A Sensor with Analog and Digital Pixels\\ in 0.15~$\mu$m SOI Technology}
%
%

\author{Marco~Battaglia,
        Dario~Bisello,
	Devis~Contarato,
        Peter~Denes,\\
        Piero~Giubilato,
        Lindsay~E.~Glesener,
	Serena~Mattiazzo,
	Chinh~Vu
\thanks{This work was supported by the Director, Office of Science, of the
U.S. Department of Energy under Contract No.DE-AC02-05CH11231.}
\thanks{M.~Battaglia and L.E.~Glesener are with the Department of Physics, University of 
California and the Lawrence Berkeley National Laboratory, Berkeley, CA 94720, USA.}%
\thanks{D.~Contarato is with the Lawrence Berkeley National Laboratory, 
Berkeley, CA 94720, USA.}%
\thanks{D.~Bisello and S.~Mattiazzo are with the Istituto Nazionale Fisica 
Nucleare, Sezione di Padova, Italy.}%
\thanks{P.~Giubilato is with the Lawrence Berkeley National Laboratory, 
Berkeley, CA 94720, USA and the Universita' degli Studi di Padova, Dipartimento di Fisica, 
Padova, Italy.}%
\thanks{P.~Denes and C.~Vu are with the Lawrence Berkeley National Laboratory, 
Berkeley, CA 94720, USA.}%
}

\maketitle
\pagestyle{empty}
\thispagestyle{empty}

\begin{abstract}
A monolithic pixel sensor in deep-submicron Silicon-On-Insulator (SOI) 
CMOS technology has been designed, manufactured and characterised.
This technology is of significant interest for applications in particle 
tracking and imaging. The prototype chip features pixels of 10~$\mu$m
pitch arrayed in two analog sections and one digital section with a comparator and 
a latch integrated in each pixel. The prototype response has been tested with 
infrared lasers and with the 1.35~GeV electron beam extracted from the 
injection booster at the LBNL Advanced Light Source. Results from irradiation tests 
with low energy protons and neutrons performed at the LBNL 88-inch Cyclotron are 
also presented.
\end{abstract}


\section{Introduction}

\IEEEPARstart{S}{ilicon} on insulator (SOI) technology allows to fabricate
CMOS circuits on a thin Si layer, electrically insulated from the rest
of the wafer. The isolation of the electronics from the detector volume
offers clear advantages for designing pixel sensors for particle
detection, compared to bulk CMOS active pixel sensors, realised in standard CMOS
process. Both nMOS and pMOS transistors can be built, without disturbing
the charge collection, and the detector wafer can be reversely biased, 
thus improving the efficiency of charge carriers collection. There have been
earlier attempts to develop SOI pixel sensors for charged particle detection
with a high resistivity bottom wafer. A proof of principle of the concept was
obtained using a 3~$\mu$m process at IET, Poland~\cite{soi-iet1,soi-iet2,soi-iet3}.
The availability of the 0.15~$\mu$m FD-SOI process by OKI Electric Industry 
Co.\ Ltd.\, Japan,  in cooperation 
with KEK, Japan, has opened up new possibilities for SOI pixel sensors with 
a pixel pitch, which is small enough to satisfy the requirements for the next
generation of particle physics collider experiments, as well as for imaging. 
As shown in Figure~\ref{fig:soi}, the OKI SOI process consists of an $n$-type detector 
wafer, with $n$- and $p$-type implants. Transistors are di-electrically isolated from 
each other, and separated from the substrate by a buried oxide layer (BOX).  
The silicon film on which the transistors are grown is sufficiently thin (40~nm) 
and lightly doped so that it can be fully depleted at typical operating voltages. 
A chip based on this process has already been designed at KEK, Japan and 
successfully tested with an IR laser beam~\cite{soi-oki1,soi-oki2}.

We designed and submitted a monolithic pixel sensor chip, with 
10$\times$10~$\mu$m$^2$ pixels, for charged particle detection. The chip has 
both analog and digital pixel cells. This paper presents the chip design and the 
results of its characterisation, obtained with laser beams in the lab 
and with the 1.35~GeV electrons beam from the LBNL Advanced Light Source 
(ALS) booster at LBNL.
\section{Prototype chip design}
\begin{figure}[b]
\begin{center}
\epsfig{file=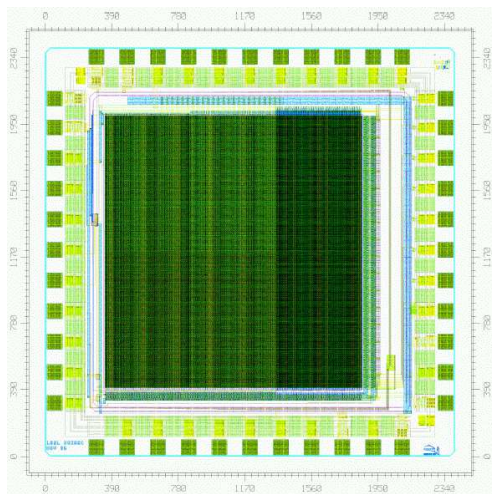,width=6.5cm}
\end{center}
\caption[]{Layout of the first LBNL SOI prototype pixel chip.}
\label{fig:chip_layout}
\end{figure}

\begin{figure}[th!]
\begin{center}
\epsfig{file=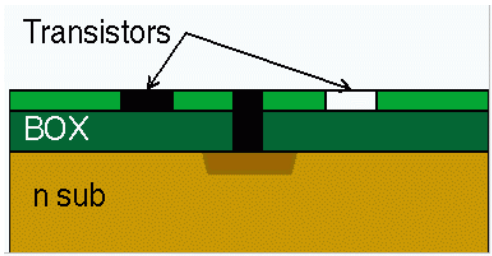,width=6.5cm}
\end{center}
\caption[]{Simplified cross section of a chip in the OKI SOI technology.}
\label{fig:soi}
\end{figure}

The chip consists of a 350~$\mu$m thick high-resistivity $n$-type substrate, 
with the CMOS circuitry implanted on a 40~nm thin Si layer on top of a 200~nm 
thick buried oxide (see Figure~\ref{fig:soi}). 

\begin{figure}[h!]
\begin{center}
\epsfig{file=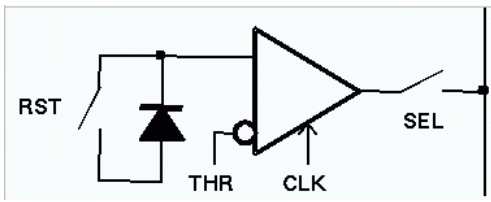,width=6.5cm}
\end{center}
\caption[]{Schematics of the clocked digital pixel.}
\label{fig:digpixel}
\end{figure}

The chip features an array of 160$\times$150 pixels on a 10~$\mu$m pitch. 
The OKI SOI technology includes
thin-oxide 1.0~V transistors and  thick-oxide 1.8~V transistors. The left-most
50 columns are simple analog pixels with 1.8~V transistors in a 
3T-like (complementary switches) architecture. 
The central 50 columns implement 
1.0~V analog pixels, and the right-most 50 columns are clocked, digital pixels. 
The digital pixel is based on a clocked comparator.  When the pixel is reset, the same 
reset being used for the comparator, charge is accumulated on the diode capacitance.  
As the row is selected for readout, the comparator is clocked, and the pixel is 
considered to be hit if the voltage on the diode is greater than a common threshold
(see Figure~\ref{fig:digpixel}). In order to avoid static power dissipation, no 
amplifier is present in the digital pixels,

The 1.0~V pixels have been found
to have significantly higher leakage currents than the 1.8~V pixels, which adversely
affect their signal-to-noise (S/N) ratio. Each sector is divided in two subsections 
with 1$\times$1~$\mu$m$^2$ and 4$\times$4~$\mu$m$^2$ charge collecting
diodes. Each 8000-pixel analog section is read out independently using a 14-bit ADC.
A Xilinx FPGA controls all pixel clocks and resets. The pixels are clocked at
6.25~MHz corresponding to an integration time of 1.382~ms. Correlated Double Sampling 
(CDS) is performed by acquiring two frames of data with no pixel reset between the 
readings and subtracting the first frame from the second.

\begin{figure}[h!]
\begin{center}
\epsfig{file=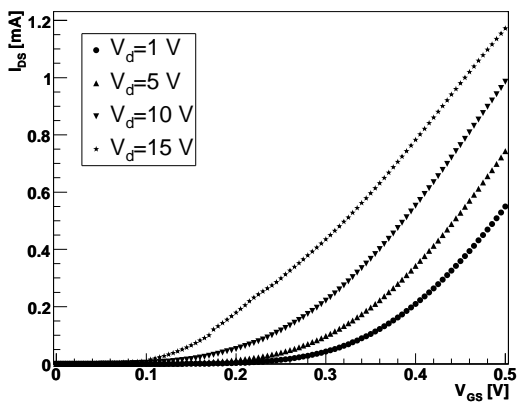,width=7.0cm}
\end{center}
\caption[]{Shift of the input characteristics $I_{DS}(V_{GS})$ as a function of the substrate 
bias $V_d$ for a 1.0~V nMOS test transistor with $W/L$=50/0.3.}
\label{fig:threshold_shift}
\end{figure}

A potential limitation of the SOI technology comes from the transistor back-gating
effect. The reverse bias, $V_d$, applied to the silicon substrate increases the 
potential at the surface, so that the buried oxide acts as a second gate for 
the CMOS circuitry on top, typically causing a shift in the transistor thresholds
for increasing depletion voltages. The effect was evaluated on single transistor 
test structures implemented at the chip periphery. These include complementary 
$p$-type and $n$-type MOSFETs, all with $W$~= 50~$\mu$m and
$L$~= 0.3~$\mu$m, with different types of body contacts (floating, source-tied and
gate-tied). Figure~\ref{fig:threshold_shift} shows the $I_{DS}(V_{GS})$ characteristics
measured on one $n$-type test MOSFET for substrate voltages, $V_d$, up to 15~V. The 
transistor threshold voltage, extracted from the $\sqrt{I_{DS}(V_{GS})}$ characteristics, 
shifts from $V_{T}$~= 0.24~V at $V_d$~= 1~V to $V_T$~= 0.07~V at $V_d$~= 15~V, consistent 
with an increased back-gating effect. A similar effect is also measured on $p$-type test 
MOSFETs.

\begin{figure}[t]
\begin{center}
\epsfig{file=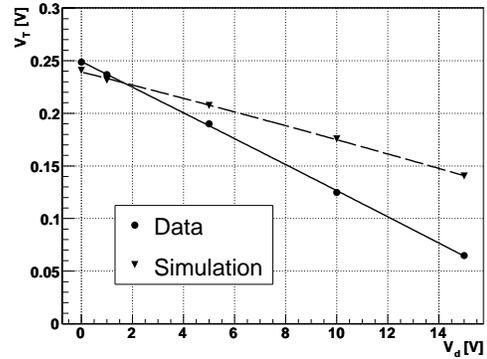,width=7.0cm}
\end{center}
\caption[]{Transistor threshold voltage, $V_T$, as a function of the substrate bias, $V_d$, 
measured for a 1.0~V test transistor (points) compared with that predicted with the 
device simulation (triangles).}
\label{fig:comp}
\end{figure}

The back-gating effect in the chip active area is investigated using TCAD simulations 
implemented with the Synopsys {\tt Taurus Device} package. A 2-dimensional model of a 5 
pixel array is used, including a realistic geometry of the substrate contact region 
at the chip edges. 
The $n$-type substrate is modelled with a constant doping level 
of 6$\times$10$^{12}$~cm$^{-3}$,
while the pixel diodes are modelled with shallow (500~nm deep) $p$-type implants
with a peak doping concentration of 1$\times$10$^{20}$~cm$^{-3}$ at the Si surface,
decreasing along a Gaussian profile towards the Si substrate.   
\begin{figure}[ht!]
\begin{center}
\epsfig{file=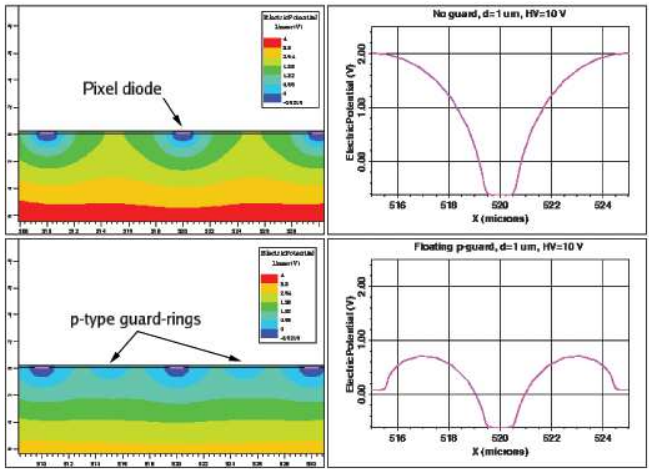,width=\columnwidth}
\end{center}
\caption[]{Simulation of the electrostatic potential in the silicon substrate
with depletion voltage $V_d$~= 10~V for pixels without (top) and with 
(bottom) floating $p$-type guard-ring. The plots on the right show the potential
along a section at the surface between the silicon substrate and the buried oxide.
All dimensions are in $\mu$m.}
\label{fig:tcad_sim}
\end{figure}
The characteristics of single transistor at the chip periphery have been 
simulated as a function of the substrate bias. Figure~\ref{fig:comp} compares 
the simulated thresholds for a n-MOSFET with the measurements performed on a 
test transistor, which show a qualitative agreement, within the uncertainty in 
reproducing the actual process parameters. The electrostatic
potential at the interface between the buried oxide and the silicon substrate is 
simulated as a function of the substrate bias for different pixel layouts,
varying the size of the charge collecting diodes and evaluating the effect
of different guard-ring configurations. This study indicates that the most effective 
design has a floating $p$-type guard-ring around each pixel.
Figure~\ref{fig:tcad_sim} shows the guard-rings effect on the field in the area between 
diode implants, which limits the potential back-gating. According to this results, a 
floating $p$-type guard-ring has  been implemented around each pixel in the active area 
of the test chip. A series of floating and grounded guard-rings has also been 
implemented around the pixel matrix and around the peripheral I/O electronics.

\section{Tests with infrared lasers}
\begin{figure}[hb!]
\begin{center}
\epsfig{file=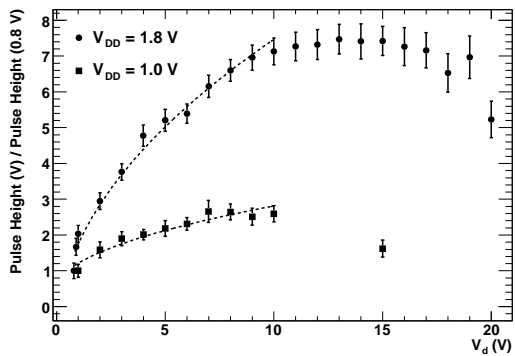,width=7.0cm}
\end{center}
\caption[]{Cluster pulse height normalised to that measured at 
$V_d$ = 0.8~V for a focused 1060~nm laser spot as a function of the 
substrate bias, $V_d$.}
\label{fig:laser}
\end{figure}

The response of the analog sections has been tested with a 1060~nm IR laser,
for different $V_d$ values.  The laser was focused to a $\simeq$~20~$\mu$m spot
and pulsed for 30~$\mu$s between successive readings. We measured the signal
pulse height in a 5$\times$5 matrix,  centred around the laser spot centre.
The measured signal increases as $\sqrt{V_d}$, as expected from the corresponding 
increase of the depletion region, until $V_d \simeq$ 9~V, where it saturates, 
to decrease for $V_d \ge$ 15~V (see Figure\ref{fig:laser}). We interpret this 
effect as due to transistor back-gating, affecting the 1.0~V transistor pixels
at lower $V_d$ values compared to the 1.8~V transistor pixels.

\begin{figure}[h!]
\begin{center}
\epsfig{file=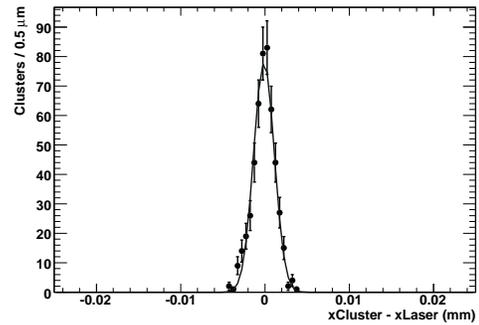,width=7.0cm}
\end{center}
\caption{Distribution of the differences between the position of a 5~$\mu$m laser spot 
and the reconstructed cluster position in the SOI analog pixels. The measurement has 
been performed with a S/N of 15 and a depletion voltage $V_d$ = 7~V. The r.m.s. of the 
fitted Gaussian function is 1.2~$\mu$m.}
\label{fig:res_sn15}
\end{figure}

\begin{figure}[h!]
\begin{center}
\epsfig{file=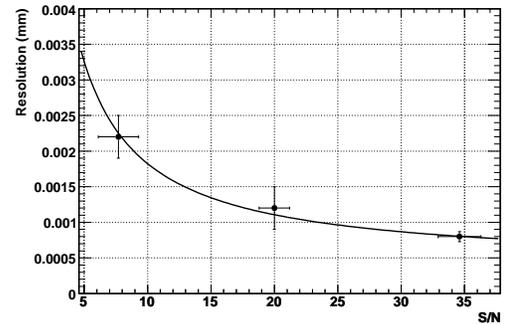,width=7.0cm}
\end{center}
\caption{Single point resolution as a function of the S/N ratio obtained from
scans of the analog pixels performed with a 1060~nm laser focused to a 
5~$\mu$m spot. The overlayed function shows the expected (S/N)$^{-1}$ scaling.}
\label{fig:res}
\end{figure}

The spatial resolution of the analog pixels is determined using a
1060~nm pulsed laser beam focused to a $\simeq$~5~$\mu$m Gaussian spot.
We performed pixel scans by shifting the laser spot along single pixel rows 
in steps of 1~$\mu$m using a stepping motor with a positioning accuracy of 
$\simeq$~0.1~$\mu$m. For each position we record 500 events. We reconstruct the 
hit position from the centre of gravity of the reconstructed cluster charge 
and study the linearity between the average cluster position and the spot position, 
obtained from the reading of the stepping motor linear encoder. The resolution is 
extracted from the spread of the reconstructed cluster position for sets of events 
taken at each point in the scan (see Figure~\ref{fig:res_sn15}). The laser intensity 
is varied to obtain different S/N values, from values below that observed for 1.35~GeV 
electron signals up to S/N~$\simeq$~35. Pixels with 10~$\mu$m pitch have a single 
point resolution of 1~$\mu$m for a S/N ratio of 20 or larger and the measured 
resolution scales as the inverse of S/N, as expected (see Figure~\ref{fig:res}).
Similar results are observed also for $V_d$ = 10~V.

\section{Tests with high-energy particle beams}
\begin{table*}[t!]
\caption{Summary of ALS beam test results on the digital pixels.
The average number of clusters per ALS spill recorded with beam on and
beam off and the average pixel multiplicity in a cluster are given for 
different values of $V_d$. The beam intensity was not constant for the 
various runs.}
\label{tab:als_digi}
\begin{center}
\begin{tabular}{|c|c|c|c|c|c|c|c|c|c|c|}
\hline  \textbf{$V_d$} & \textbf{$\frac{Nb. Clusters}{Spill}$} & \textbf{$\frac{Nb. Clusters}{Spill}$} & \textbf{$<$Nb Pixels$>$} \\ 
\textbf{(V)}           & \textbf{beam on}                      & \textbf{beam off}                     &\textbf{in Cluster}       \\ 
\hline
20                     & 3.62                                  & 0.02                                  & 1.78                     \\ 
25                     & 5.81                                  & 0.03                                  & 1.32                     \\
30                     & 8.31                                  & 0.03                                  & 1.26                     \\
35                     & 1.60                                  & 0.02                                  & 1.14                     \\
\hline
\end{tabular}
\end{center}
\end{table*}

The pixel chip has been tested on the 1.35~GeV electron beam-line at LBNL
Advanced Light Source (ALS). The readout sequence of a reference frame followed 
by the signal frame is synchronised with the 1~Hz booster extraction cycle 
so that the beam spill hits the detector just before the second frame is 
read out. The temperature is kept constant during
operation at $\simeq$~23$^{\circ}$C by forced airflow.
Data are processed on-line by a {\tt LabView}-based program, which performs 
correlated double sampling, pedestal subtraction and noise computation. 
The data are converted in the {\tt lcio} format and the offline analysis
is performed using a set of dedicated processors developed in the 
{\tt Marlin} framework~\cite{Gaede:2006pj}. Each event is scanned for
seed pixels with pulse height above a signal-to-noise threshold of 4.5.
Noisy pixels are flagged and masked. Seeds are sorted according to their
pulse heights and the neighbouring pixels in a 5$\times$5 matrix with
S/N$>$2.5 are added to the cluster. Clusters are not allowed to overlap 
and we require that pixels associated to a cluster are not interleaved by 
any pixels below the neighbour threshold. 

All the sections are functional. Results for two of the analog sections,
with $V_{DD}$~= 1.0~V and 1.8~V transistors respectively, each consisting of
a 0.4$\times$0.8~mm$^2$ active region, have already been 
presented~\cite{Battaglia:2007eq}. 
Data have been taken at different depletion voltages, $V_d$, from 1~V up
to 15~V for the analog sections, and up to 30~V for the digital section.
The depletion voltages used correspond to an estimated depletion thickness
of 8~$\mu$m to 80~$\mu$m for $V_d$ = 1~V to 30~V. Only a small background is 
found, arising from noisy pixels which survive the bad pixel cut and the cluster 
quality criteria adopted in the data analysis. The pixel multiplicity in a 
cluster decreases with increasing depletion voltage, while the cluster 
pulse height increases for $V_d$ up to 10~V. At $V_d \simeq$~15~V the cluster 
signal and the efficiency of the chip decreases, similarly to the trend observed 
in the laser test. A good signal-to-noise ratio up to 15 as measured with 
the analog section with 1.8~V transistors for 5~V$\le V_d \le$15~V.

\begin{figure}[ht]
\begin{center}
\epsfig{file=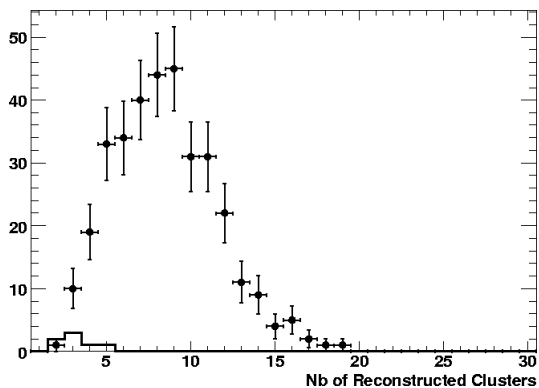,width=7.0cm}
\end{center}
\caption[]{Hit multiplicity for events taken with the 1.35~GeV $e^-$ beam on 
digital pixels at a depletion voltage of 30~V (markers with error bars). 
The distribution of fake hits reconstructed in the absence of beam 
(continuous line) is also shown.}
\label{fig:als_digi}
\end{figure}

The digital section of the chip is also found to be functional. Beam signals 
are observed by applying depletion voltages in excess of 20~V, and up to 30~V. 
This can be explained by considering that the analog threshold of the
in-pixel comparators is also affected by back-gating, but the digital circuitry
in each pixel is only active when triggered, i.e.\ for times much shorter compared 
to that of the analog pixels. As larger substrate voltages are needed to 
obtain signals large enough to be above threshold, the two effects combine 
giving the best particle detection capabilities for 20~V~$\le V_d \le$~30~V. 
Figure~\ref{fig:als_digi} shows the hit multiplicity observed in the digital
pixels for events taken with and without beam, for $V_d$ = 30~V. A clear excess 
of hits can be seen in the presence of beam. The average number of pixels in a
cluster decreases from 1.8 to 1.3 for $V_d$ increasing from 20~V
to 30~V (see Table~\ref{tab:als_digi}). 

\section{Radiation hardness tests}
\begin{figure}[ht!]
\begin{center}
\epsfig{file=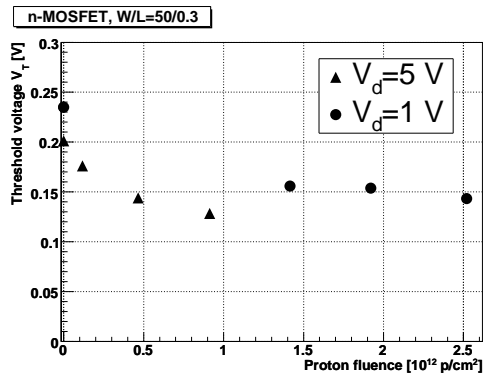,width=7.0cm}
\end{center}
\caption{Threshold voltage for a 1.0~V nMOS test transistor with $W/L$ = 50/0.3 as a function of the 30~MeV proton
fluence for two depletion voltage values $V_d$ = 1~V and $V_d$ = 5~V. The measurements were performed on the same 
test transistor soon after each irradiation step.}
\label{fig:irrad}
\end{figure}

Irradiation tests have been performed at the BASE Facility of the LBNL 88-inch Cyclotron~\cite{cyclotron}.
A first test has been performed with 30~MeV protons on single transistors. The chip is mounted on the 
beam line, behind a 1-inch diameter collimator, and the terminals of two test transistors (one $p$-MOSFET 
and one $n$-MOSFET) are connected to a semiconductor parameter analyser so that the transistor characteristics
could be measured in-between irradiation steps. During the irradiation steps, the transistor terminals
are kept grounded. The irradiation was performed with a flux of $\simeq$~6$\times$10$^7$~p/cm$^2$s,
up to a total fluence of 2.5$\times$10$^{12}$~p/cm$^2$, corresponding to a total dose of $\simeq$~600~kRad.
Figure~\ref{fig:irrad} shows the variation in the threshold voltage for the nMOS test transistor as a
function of the proton fluence. An initial substrate voltage $V_d$ = 5~V was used, but after a fluence of
about 1$\times$10$^{12}$~p/cm$^2$ the transistor characteristics could not be properly measured, and a
reduced substrate bias of $V_d$ = 1~V needed to be apply in order to recover the transistor characteristics.
We interpret this effect as due to radiation-induced charge build-up in the buried oxide which effectively
increases back-gating. The total threshold variation is indeed significant ($\sim$~100~mV) also for a low
substrate bias (i.e. $V_d$ = 1~V). The effect is much larger than what would be expected at such doses
from radiation damage in the transistor thin gate oxide. Similar results are obtained for the pMOSFET 
characteristics.

A second test has been performed on a fully functional chip exposed to a beam of 1-20~MeV neutrons,
up to a total fluence of 1$\times$10$^{11}$~n/cm$^2$. No change in the noise of the analog sections 
is observable, and no significant variation in the test transistor characteristics is detected. 

\section{Second prototype chip design}
A second prototype chip has been designed and submitted for fabrication in the 0.20~$\mu$m
SOI process from OKI. This process has been is optimised for low leakage current. The chip is 
a 5$\times$5~mm$^2$ prototype with an active area of 3.5$\times$3.5~mm$^2$ in which
168$\times$172 pixels are arrayed with a 20~$\mu$m pitch. The pixel matrix is subdivided into
a 40$\times$172 pixel section with a simple, analog 3-transistor architecture, mostly
intended for technology evaluation, and a 128$\times$172 pixel main section with a 
second-generation digital pixel cell. Each pixel has two capacitors for in-pixel CDS, and a
digital latch is triggered by a clocked comparator with a current threshold. The chip design 
is optimised for readout up to a 50~MHz clock frequency, and the digital section
has multiple parallel outputs to further improve the frame rate.

\section{Summary and conclusion}
A pixel chip including both analog and digital pixel cells has been designed and 
produced in OKI 0.15~$\mu$m SOI technology. The response of the chip 
has been studied using both focused laser beams and the 1.35~GeV $e^-$ beam at the 
LBNL ALS. Both the analog and digital pixels are functional.
In particular, on the analog pixels, a single point resolution of $\simeq$~1~$\mu$m 
is estimated from a focused laser scan. The results of total ionising dose tests 
performed on single transistors hint at the observation of trapped charge build-up 
in the buried oxide, which enhances the CMOS electronics back-gate effect. 
No sensitivity to non-ionising
dose has been observed up to a fluence of 10$^{11}$~n/cm$^2$. These results are 
very encouraging for the further development of monolithic pixel sensors in SOI technology. 
The SOI technology is of great interest for its potential to implement complex readout 
architectures combined with a high-resistivity, depleted substrate. Ensuring faster 
charge collection and larger signals SOI pixel sensors are an attractive option for 
application in particle tracking at future colliders as well as non-destructive beam 
monitoring and fast imaging of low energy radiation.

\vspace*{-0.1cm}

\section*{Acknowledgements}

\vspace*{-0.1cm}

This work was supported by the Director, Office of Science, of the
U.S. Department of Energy under Contract No.DE-AC02-05CH11231.
We are indebted to the ALS staff for their assistance and the
excellent performance of the machine.

\end{document}